\newif\ifCOMMENT
\newif\ifTODO
\documentclass[12pt]{article}
\usepackage{amsmath,bm}
\usepackage[dvipdfm]{graphicx,color}
\usepackage{cite}
\def\a{\alpha}
\def\b{\beta}
\def\g{\gamma}
\def\d{\delta}
\def\D{\Delta}
\def\e{\epsilon}
\def\k{\kappa}
\def\th{\theta}
\def\lam{\lambda}
\def\del{\partial}
\def\half{\frac{1}{2}}
\newcommand{\ltsim}{\mathrel{\hbox{\raise0.2ex
\hbox{$<$}\kern-0.75em\raise-0.9ex\hbox{$\sim$}}}}
\newcommand{\C}[1]{\ifCOMMENT{\color{red} #1}\else\fi}

\begin{document}
\flushright{KYUSHU-HET 73\\September 2004}
\bigskip
\begin{center}
{\Large{\bf The Higgs sector in the Next-to-MSSM}}\\[10mm]
Koichi \textsc{Funakubo}$^{a,}$\footnote{E-mail: funakubo@cc.saga-u.ac.jp} 
and Shuichiro \textsc{Tao}$^{b,}$\footnote{E-mail: tao@higgs.phys.kyushu-u.ac.jp}\\[5mm]
{\it $^a$Department of Physics, Saga University, Saga 840-8502, Japan \\
$^b$Department of Physics, Kyushu University, Fukuoka 812-8581, Japan}
\end{center}
\abstract{
We study the Higgs sector in the Next-to-Minimal Supersymmetric Standard Model 
with and without explicit CP violation, focusing on the case of weak scale expectation
value of the singlet field.
We scan a wide range of the parameter space to find out allowed regions by requiring
that the electroweak vacuum be the global minimum of the effective potential and that
the neutral Higgs bosons with moderate gauge coupling be heavier than the lower bound
on the Higgs boson in the standard model.
Among the allowed parameters, some sets admit the situation in which the light Higgs bosons couple
with the $Z$ boson too weak to be found in present collider experiments.
For such parameter sets, 
we find an upper bound on the charged Higgs mass which is reachable in LHC.}
%
\section{Introduction}
\label{sec:introduction}
\C{Higgs mass bound}
The search for the Higgs boson is one of the most important issue 
of high-energy particle physics because the Higgs boson 
is the only unobserved particle in the Minimal Standard Model (MSM).
The results of the LEP~2 experiment place the limit on the MSM-Higgs mass: 
$m_h >114.4 {\rm GeV}$ at $95\%$ CL\cite{barate03,pdg}.
Although there are some theoretical restrictions to the Higggs mass, it cannot
be predicted in the MSM framework, since the Higgs self-coupling is a free parameter.
Supersymmetric extensions of the MSM, which are motivated to solve the hierarchy problem,
limit the range of the Higgs mass because of the self-coupling given by the gauge couplings.
Among such extensions, the Minimal Supersymmetric Standard Model (MSSM) has been
well studied and is known to give the upper mass bound of the lightest Higgs boson, $m_h\le m_Z$ 
at the tree level.
This bound seems somewhat severe but it is modified by radiative corrections, 
shifting it to about $m_h\le 135{\rm GeV}$ at the two-loop level, mainly by the loops
of the top quark and squark\cite{higgs-mass-correction}.

\C{NMSSM}
The MSSM contains a $\mu$-parameter in the superpotential. It enters the Higgs potential with
the soft scalar masses to determine the vacuum expectation value (VEV) of the Higgs fields.
Then $\mu$ must take a value of weak scale, which is much smaller than the GUT scale or Planck scale.
There is no a priori reason for $\mu$ to have such a small value.
One solution of this so-called $\mu$-problem is to substitute a VEV of an extra gauge-singlet field for 
the parameter $\mu$.
The NMSSM is among such models which have a gauge-singlet Higgs superfield $N$. 
The superpotential of the model contains
\begin{equation}
 W = -\lam NH_dH_u -\frac{1}{3}\k N^3,
\end{equation}
in addition to the MSSM terms with $\mu=0$\cite{fayet75}. 
We adopt the $Z_{3}$-symmetric version of the superpotential so that it does not 
contain any dimensional coupling.
The $\mu$ parameter of the MSSM is generated as $\mu = \lam\left<N\right>$.
The NMSSM behaves like the MSSM in the limit of $\langle N\rangle\gg v$ with $\lam\langle N\rangle$ 
and $\k\langle N\rangle$ fixed, for which the singlet decouples. 
Here $v=\sqrt{\langle H_d\rangle^2+\langle H_u\rangle^2}$ is the VEV of the Higgs doublets.
Since the superpotential has no dimensional parameters, the VEVs of the Higgs fields are determined
by the mass parameters in the soft-supersymmetry-breaking terms together with the couplings.
If all the mass parameters in the Higgs potential are of weak scale and all the couplings have moderate 
values about $0.1-1$, it is natural for the singlet to acquire a VEV of the same order as $v$.
Then we expect new features which are absent from the MSSM, and our main concern is in the Higgs
sector of this case.

\C{NMSSM spectrum, CP cons.}
The NMSSM contains three CP-even neutral Higgs bosons $(S_1, S_2, S_3)$, two CP-odd bosons
$(A_1, A_2)$ and a pair of charged bosons $H^\pm$ in the CP-conserving case.
The spectrum of the CP-conserving model has been studied in \cite{gunion86,ellis89,miller03}.
In contrast to the MSSM, the three CP-even scalars can mix up to form mass eigenstates of small mass
with very small gauge coupling\cite{miller03}. Such a light Higgs boson cannot be produced
at lepton colliders so that it is not excluded even if its mass is smaller than $114\mbox{GeV}$.
We shall refer to this situation as {\it light Higgs} scenario.
As we see below, such a light Higgs situation is realized for weak scale $\langle N\rangle$ and small $\kappa$.
A similar situation has been observed in the MSSM, when a large mixing among CP eigenstates is 
caused by the CP-violation in the squark sector, which is characterized by the imaginary part of
the product of $\mu$ and the $A$-term\cite{squark-phase-mssm}.
Then the gauge coupling of the lightest scalar is so small that it can escape the lower bound on the
Higgs mass\cite{CPXhiggs}.
While the same mixing is also expected in the NMSSM, it contains another source of CP violation
in the tree-level Higgs sector.
Such a CP violation has been studied in several cases where it is caused spontaneously\cite{nmssm-spontaneous-cpv} 
and explicitly\cite{nmssm-explicit-cpv} in some special situations. 
We give the one-loop formulation with all possible CP phases including squark sector
in a manner independent of phase convention. 
In this formulation, one can easily arrange the phases in such a way that
the phase relevant to the neutron Electric Dipole Moment (nEDM) is suppressed 
while retaining those which affect the mixing of the Higgs bosons. 
We investigated the mass spectrum and couplings in the presence of such a CP phase.

Another aspect of the NMSSM Higgs sector is that the Higgs potential contains cubic terms including the singlet field. 
Although these terms must be constrained not to generate the global minimum of the effective potential
different from the electroweak vacuum, they are expected to make stronger the first-order phase transition
at high temperature.
In this sense, this model is more suited for electroweak baryogenesis than the MSSM, which requires
a light stop with mass less than the top quark mass for the strongly first-order phase transition\cite{nmssm-ewpt}.
We pointed out that the CP violation in the squark sector weakens the electroweak phase transition (EWPT) caused
by a light stop in the MSSM\cite{funakubo03}.
We expect that, in contrast to the MSSM, the CP violation in the Higgs sector of the NMSSM will not
weakens the EWPT, while supplying sufficient CP violation to generate chiral charge flux, which is the source
of baryon asymmetry.
The effects of this CP violation on the phase transition will be discussed in our forth-coming paper.

\C{Sections}
This paper is organized as follows.
In Section~\ref{sec:tree-level}, we analyze the NMSSM Higgs sector at the tree level 
and explain how to obtain restrictions on the parameters in the model. There
we derive the upper and lower bounds on the mass of the charged Higgs, which
are trivial in the MSSM limit but important in the case of weak scale $\langle N\rangle$.
In Section~\ref{sec:one-loop}, we show the one-loop formulas for the mass-squared matrix of
the neutral Higgs bosons and a mass of the charged Higgs boson.
Section~\ref{sec:parameter-search} is devoted to the numerical results for the parameter search.
The spectrum condition divides the allowed parameter sets into two classes.
The first one is the MSSM-like allowed parameter sets where all the Higgs bosons are heavier than the $114\:{\rm GeV}$, and 
the second one corresponds to the light Higgs scenario.
The study of the CP violation is described in Section~\ref{sec:cp-violation}.
The formulas used to define the effective potential and to calculate the mass matrix are
summarized in Appendices.

\section{Tree-level Higgs sector}
\label{sec:tree-level}
\subsection{Higgs potential}
\C{Superpotential}
In this Section we analyze the tree-level Higgs sector.
The NMSSM has the superpotential with the singlet superfield $N$\cite{miller03}, 
\begin{equation}
 W=f_dH_dQD^c-f_uH_uQU^c
  -\lam NH_dH_u-\frac{\k}{3}N^3,
 \label{eq:superpotential}
\end{equation}
where $Q$, $D^c$ and $U^c$ denote chiral superfields containing quarks, and 
$H_d$ and $H_u$ 
contain the Higgs doublets required in the MSSM.
The couplings $\lam$ and $\k$ for the singlet are in general complex numbers.
We consider $Z_{3}$-symmetric version of the superpotential, 
so it does not contain any dimensional coupling.

\C{soft-term}
In addition to the supersymmetric Lagrangian, the low-energy NMSSM 
contains the soft-SUSY-breaking terms, 
\begin{align}
  \mathcal{L}_{\rm soft}= &-m_1^2\Phi_d^\dag\Phi_d
  -m_2^2\Phi_u^\dag\Phi_u-m_n^2n^*n 
  -m_{\tilde{q}}^2\tilde{q}_L^\dag\tilde{q}_L
  -m_{\tilde{d}}^2\tilde{d}_R^\dag\tilde{d}_R
  -m_{\tilde{u}}^2\tilde{u}_R^\dag\tilde{u}_R\notag\\
 &-\left\{(f_dA_d)\Phi_d\tilde{q}_L\tilde{d}_R^*
  -(f_uA_u)\Phi_u\tilde{q}_L\tilde{u}_R^*
  -\lam A_\lam n\Phi_d\Phi_u+{\rm h.c.}\right\}\notag\\
 &-(m_n'n^2+\frac{1}{3}\k A_\k n^3+{\rm h.c.}),
  \label{eq:susyxsoftterm}
\end{align}
where $\tilde{q}_L$, $\tilde{d}_R$ and $\tilde{u}_R$ are the
squark fields and $\Phi_d$, $\Phi_u$ and $n$  are the Higgs fields.
Although the $n^2$-term breaks the global $Z_3$-symmetry, the $n^2$-term does not exist in the simple 
supergravity model. Hence we shall not include the $n^2$-term in the following.
We assume that  all the dimensional parameters in $\mathcal{L}_{\rm soft}$
have their values near weak scale.

\C{$Z_{3}$-symmetry}
In the limit of $\kappa=0$, the global $Z_3$-symmetry is elevated to $U(1)$ PQ symmetry 
($n\to e^{i\a}n$ and $\Phi_{u}\to e^{-i\a}\Phi_{u}$).  Then the pseudoscalar component of the 
singlet becomes a Nambu-Goldstone (NG) mode when the singlet acquires a VEV.
At small $\k$, the PQ symmetry is slightly broken and relatively light axion is expected\cite{axion}.
The spontaneous breakdown of the global $Z_3$-symmetry causes the domain wall problem\cite{domain-wall}.
If we introduced $Z_3$-breaking linear or bilinear terms into the superpotential, 
this problem could be solved\cite{menon04}.
It inevitably spoils the nature of the superpotential with no dimensional parameter as the countervalue.
So we assume that this symmetry is broken by some higher dimensional operator which becomes irrelevant
in low energy physics, at the early stage of the universe far before the EWPT.

The tree-level Higgs potential is composed of three parts, $V=V_F+V_D+V_{\rm soft}$;
\begin{align}
  &V_F=|\lam n|^2(\Phi_d^\dag\Phi_d+\Phi_u^\dag\Phi_u)
  +|\e_{ij}\lam\Phi_d^i\Phi_u^j+\k n^2|^2,
  \label{eq:vfhiggs}\\
  &V_D=\frac{g_2^2+g_1^2}{8}(\Phi_d^\dag\Phi_d-\Phi_u^\dag\Phi_u)^2
  +\frac{g_2^2}{2}(\Phi_d^\dag\Phi_u)(\Phi_u^\dag\Phi_d),
  \label{eq:vdhiggs}\\
  &V_{\rm soft}=m_1^2\Phi_d^\dag\Phi_d+m_2^2\Phi_u^\dag\Phi_u+m_N^2|n|^2
  -(\e_{ij}\lam A_\lam n\Phi_d^i\Phi_u^j+\frac{1}{3}\k A_\k n^3+{\rm h.c.}).
  \label{eq:vsofthiggs}
\end{align}
Here we expand this potential around the VEVs which are represented by 
$v_d$, $v_u$, $v_n$ and the phases $\th$ and $\varphi$. 
The parametrization of the scalar fields are as follows,
\begin{equation}
 \Phi_d =
 \left(\begin{array}{c}
  \frac{1}{\sqrt{2}}(v_d+h_d+ia_d)\\ \phi_d^-
 \end{array}\right),\quad
 \Phi_u =
 e^{i\th}\left(\begin{array}{c}
  \phi_u^+\\ \frac{1}{\sqrt{2}}(v_u+h_u+ia_u)
 \end{array}\right),
\end{equation}
\begin{equation}
 n = \frac{1}{\sqrt{2}}e^{i\varphi}(v_n+h_n+ia_n).
\end{equation}

\C{Tadpole condition}
The condition for the scalar potential to have an extremum at the
vacuum is that the first derivatives with respect
to the Higgs fields evaluated at the vacuum vanish:
\begin{align}
 0=&\frac{1}{v_d}\left<\frac{\del V_0}{\del h_d}\right>
 =\tilde{m}_1^2
 -R_\lam\frac{v_uv_n}{v_d}
 +\frac{g_2^2+g_1^2}{8}(v_d^2-v_u^2)
 +\frac{|\lam|^2}{2}(v_u^2+v_n^2)
 +\frac{\mathcal{R}}{2}\frac{v_uv_n^2}{v_d},
 \label{eq:tad-hd}
\\ %
 0=&\frac{1}{v_u}\left<\frac{\del V_0}{\del h_u}\right>
 =\tilde{m}_2^2
 -R_\lam\frac{v_dv_n}{v_u}
 -\frac{g_2^2+g_1^2}{8}(v_d^2-v_u^2)
 +\frac{|\lam|^2}{2}(v_d^2+v_n^2)
 +\frac{\mathcal{R}}{2}\frac{v_dv_n^2}{v_u},
 \label{eq:tad-hu}
\\ %
 0=&\frac{1}{v_n}\left<\frac{\del V_0}{\del h_N}\right>
 =\tilde{m}_N^2
 -R_\lam\frac{v_dv_u}{v_n}
 -R_\k v_n
 +\frac{|\lam|^2}{2}(v_d^2+v_u^2)
 +|\k|^2v_n^2
 +\mathcal{R}v_dv_u,
 \label{eq:tad-hn}
\\ 
 0=&\frac{1}{v_u}\left<\frac{\del V_0}{\del a_d}\right>
 =\frac{1}{v_d}\left<\frac{\del V_0}{\del a_u}\right>
 = I_\lam v_n
   -\half\mathcal{I}v_n^2,
 \label{eq:tad-a1}
\\ %
 0=&\frac{1}{v_n}\left<\frac{\del V_0}{\del a_N}\right>
 =I_\lam\frac{v_dv_u}{v_n}
 +I_\k v_n
 +\mathcal{I}v_dv_u,
 \label{eq:tad-a2}
\end{align}
where 
\begin{align}
 &\mathcal{R}={\rm Re}[\lam\k^*e^{i(\th-2\varphi)}],&
 &\mathcal{I}={\rm Im}[\lam\k^*e^{i(\th-2\varphi)}],
\label{eq:rr}\\ %
 &R_\lam=\frac{1}{\sqrt{2}}{\rm Re}[\lam A_\lam e^{i(\th+\varphi)}],&
 &I_\lam=\frac{1}{\sqrt{2}}{\rm Im}[\lam A_\lam e^{i(\th+\varphi)}],
\label{eq:rl}\\%
 &R_\k=\frac{1}{\sqrt{2}}{\rm Re}[\k A_\k e^{i3\varphi}],&
 &I_\k=\frac{1}{\sqrt{2}}{\rm Im}[\k A_\k e^{i3\varphi}],
\label{eq:rk}
\end{align}
and $\langle\cdots\rangle$ denotes the value evaluated at the vacuum.
These conditions are called \textit{tadpole conditions} in the sense that 
the conditions make the tadpole diagrams vanish, if we set the Higgs fields to their VEVs.
Note that $\mathcal{R}$ and $\mathcal{I}$ are dimensionless parameters and
$R_\lam$, $R_\k$, $I_\lam$ and $I_\k$ have dimension one.
The phases only appear through the three combinations (\ref{eq:rr})--(\ref{eq:rk}) given above.
Hence our formulation so far does not depend on the convention of the phases.
From (\ref{eq:tad-a1}) and (\ref{eq:tad-a2}), we obtain two conditions
\begin{equation}
I_\lam=\half\mathcal{I}v_n,\quad 
I_\k=-\frac{3}{2}\mathcal{I}\frac{v_dv_u}{v_n}.
\label{eq:tadpole-p}
\end{equation}
Because of the two tadpole conditions on the three CP violating parameters 
$\mathcal{I}$, $I_\lam$ and $I_\k$, only one of them is the physical one.
When we introduce complex parameters, we have to manage them to 
satisfy tadpole conditions (\ref{eq:tadpole-p}).

\subsection{The mass and couplings of the Higgs scalars}
\label{sec:mass-matrix-tree}
The mass matrix of the neutral Higgs scalars, which is defined by the second derivative of the 
Higgs potential evaluated at the vacuum,  has the following structure,
\begin{equation}
 \mathcal{M}^2 =
\left(\begin{array}{cc}
\mathcal{M}_S^2&\mathcal{M}_{SP}^2\\
(\mathcal{M}_{SP}^2)^T&\mathcal{M}_P^2
\end{array}\right),
\end{equation}
where the basis is $(\bm{h}^T\;\bm{a}^T)=(h_d\;h_u\;h_n\;a_d\;a_u\;a_n)$.
Here the block components of $\mathcal{M}^2$ are given by
\begin{equation}
 \mathcal{M}_S^2=\left(\begin{array}{ccc}
R_\lam v_n\tan\b+m_Z^2\cos^2\b&
-R_\lam v_n-m_Z^2\sin\b\cos\b&
-R_\lam v_u+|\lam|^2v_nv_d\\
-\half\mathcal{R}v_n^2\tan\b&
+\half\mathcal{R}v_n^2+|\lam|^2v_dv_u&
+\mathcal{R}v_nv_u\\[2mm]
-R_\lam v_n-m_Z^2\sin\b\cos\b&
R_\lam v_n\cot\b +m_Z^2\sin^2\b&
-R_\lam v_d+|\lam|^2v_nv_u\\
+\half\mathcal{R}v_n^2+|\lam|^2v_dv_u&
-\half\mathcal{R}v_n^2\cot\b&
+\mathcal{R}v_nv_d\\[2mm]
-R_\lam v_u+|\lam|^2v_nv_d&
-R_\lam v_d+|\lam|^2v_nv_u&
R_\lam \frac{v_dv_u}{v_n}-R_\k v_n\\
+\mathcal{R}v_nv_u&
+\mathcal{R}v_nv_d&
+2|\k|^2v_n^2
\end{array}\right),
\end{equation}
\begin{equation}
 \mathcal{M}_P^2=\left(\begin{array}{ccc}
 (R_\lam-\mathcal{R}v_n/2)v_n\tan\b&
 (R_\lam-\mathcal{R}v_n/2)v_n&
 (R_\lam+\mathcal{R}v_n)v_u\\
 (R_\lam-\mathcal{R}v_n/2)v_n&
 (R_\lam-\mathcal{R}v_n/2)v_n\cot\b&
 (R_\lam+\mathcal{R}v_n)v_d\\
 (R_\lam+\mathcal{R}v_n)v_u&
 (R_\lam+\mathcal{R}v_n)v_d&
 R_\lam\frac{v_dv_u}{v_n}+3R_\k v_n-2\mathcal{R}v_dv_u
 \end{array}\right),
\end{equation}
\begin{equation}
 \mathcal{M}_{SP}^2=\left(\begin{array}{ccc}
 0& 0&
 \frac{3}{2}\mathcal{I}v_nv_u\\
 0& 0&
 \frac{3}{2}\mathcal{I}v_nv_d\\
 -\half\mathcal{I}v_nv_u&
 -\half\mathcal{I}v_nv_d&
 -2\mathcal{I}v_dv_u
\end{array}\right),
\end{equation}
where we use the tadpole conditions (\ref{eq:tad-hd})--(\ref{eq:tad-hn}) and 
(\ref{eq:tadpole-p}) to express the scalar soft masses and $I_\lam$, $I_\k$ in terms of the other parameters.
In the following, we also adopt the usual conventions
$\tan\b=v_u/v_d$ and $v^2=v_d^2+v_u^2$.
It is worth emphasizing that, if CP violation at the tree level is turned off (i.e. $\mathcal{I}=0$),
the scalar mass matrix becomes block diagonal.

The three pseudoscalars contain one NG mode, which is isolated by the $\b$ rotation
\begin{equation}
 \bm{a}
 =U(\b)\left(\begin{array}{c}G\\\bm{a}'\end{array}\right)
 =\left(\begin{array}{ccc}
 \cos\b&\sin\b&0\\
 -\sin\b&\cos\b&0\\
 0&0&1
 \end{array}\right)
 \left(\begin{array}{c}G\\a\\a_n\end{array}\right),
\end{equation}
where $G$ is the NG mode which would be eaten up by the gauge bosons. 
After isolating the NG mode, the mass term of the neutral Higgs bosons 
is given by 
\begin{equation}
\mathcal{L}_m=-\half(\begin{array}{cc}\bm{h}^T&{\bm{a}'}^T\end{array})
\left(\begin{array}{cc}
 \mathcal{M}_S^2&{\mathcal{M}_{SP}^2}'\\({\mathcal{M}_{SP}^{2}}')^T&
 {\mathcal{M}_P^{2}}'\end{array}\right)\left(\begin{array}{c}
 \bm{h}\\\bm{a}'\end{array}\right),
\end{equation}
where
\begin{equation}
{\mathcal{M}_{SP}^{2}}'=\left(\begin{array}{cc}0&\frac{3}{2}\sin\b\\
 0&\frac{3}{2}\cos\b\\-\half&-\sin2\b\end{array}\right)
\mathcal{I}v_nv,
\label{eq:scalar-pseudoscalar-mass-matrix-revised}
\end{equation}
\begin{equation}
{\mathcal{M}_P^{2}}'=\left(\begin{array}{cc}(2R_\lam-\mathcal{R}v_n)
\frac{v_n}{\sin2\b}&(R_\lam+\mathcal{R}v_n)v\\
(R_\lam+\mathcal{R}v_n)v&\frac{R_\lam}{2}\frac{v^2}{v_n}\sin2\b+3R_\k v_n
-\mathcal{R}v^2\sin2\b\end{array}\right).
\label{eq:pseudo-scalar-mass-matrix-revised}
\end{equation}
From the equations (\ref{eq:scalar-pseudoscalar-mass-matrix-revised}) and 
(\ref{eq:pseudo-scalar-mass-matrix-revised}), 
we are convinced that one of the pseudoscalar becomes massless in the limit $\k=0$ 
($\mathcal{I}=\mathcal{R}=R_{\k}=0$). 
This holds even when we include radiative corrections.
We define ${\mathcal{M}'}^2$ as the mass matrix of the neutral Higgs bosons 
after extracting the NG mode.
Then the mass of the neutral Higgs bosons are obtained by diagonalizing 
the mass matrix ${\mathcal{M}'}^2$ by orthogonal rotation
$O^T{\mathcal{M}'}^2O={\rm diag}
(m^2_{h_1}\;m^2_{h_2}\;m^2_{h_3}\;m^2_{h_4}\;m^2_{h_5})$,
where we define the matrix $O$ in such a way that the eigenvalues satisfy
\begin{equation}
  \label{eq:mass-order}
m^2_{h_1}<m^2_{h_2}<m^2_{h_3}<m^2_{h_4}<m^2_{h_5}.
\end{equation}
Without CP violation ($\mathcal{I}=0$), the mass eigenstates are also CP eigenstates $S_i$ and $A_i$, 
where $A_i$ has vanishing $g_{VVh}$ coupling ($V$ represents the $W$ and $Z$ bosons).

\C{charged-Higgs}
Similarly, the charged Higgs mass $m_{H^\pm}$ is obtained by  the $\b$ rotation of the charged 
Higgs mass matrix,
\begin{align}
 m_{H^\pm}^2
 &=\frac{1}{\sin\b\cos\b}
  \left<\frac{\del^2V}{\del\phi_d^+\del\phi_u^-}\right>
\notag\\
 &=m_W^2-\half|\lam|^2v^2
  +(2R_\lam-\mathcal{R}v_n)\frac{v_n}{\sin2\b}.
\label{eq:charged-mass}
\end{align}
We use this equation in order to substitute $m_{H^\pm}$ for $R_\lam$ by
\begin{equation}
R_\lam =\half \hat{m}^2\frac{\sin2\b}{v_n} +\half\mathcal{R}v_n, \label{eq:R_lam}
\end{equation}
where
\begin{equation}
\hat{m}^2\equiv m_{H^\pm}^2-m_W^2+\half|\lam|^2v^2.  \label{eq:def-m-hat}
\end{equation}
For example, the CP-odd components of the mass matrix is written as
\begin{equation}
{\mathcal{M}_P^2}'=\left(\begin{array}{cc}
\hat{m}^2&
\half \hat{m}^2\frac{v}{v_n}\sin2\b+\frac{3}{2}\mathcal{R}v_nv\\
\half \hat{m}^2\frac{v}{v_n}\sin2\b+\frac{3}{2}\mathcal{R}v_nv&
\frac{1}{4}\hat{m}^2(\frac{v}{v_n}\sin2\b)^2
-\frac{3}{4}\mathcal{R}v^2\sin2\b+3\mathcal{R}_\k v_n
\end{array}\right).
\end{equation}

Now we have seven mass eigenstates, but we have no obvious bounds on the eigenvalues, like
the upper (lower) bound on the lightest (heaviest) Higgs scalar in the MSSM at the tree level.
Instead, without CP violation, the inequality 
$\det\left( \hat m^2 - \mathcal{M}_{P}^{2\prime}\right) <0$
implies that $m_{A_1}^2 < \hat m^2 < m_{A_2}^2$. It is difficult to derive such inequality for
the CP-even scalars, but in the limit of $\hat m^2\gg v_0^2, v_n^2$, we have an approximate relation
$\det\left( \hat m^2 - \mathcal{M}_{S}^2\right) \ltsim 0$, which implies that
$m_{S_1}^2 < m_{S_2}^2 < \hat m^2 < m_{S_3}^2$. These relations explain the pattern of the mass
eigenvalues in the case of heavy charged Higgs boson, with the help of ${\rm Tr}\mathcal{M}_{S}^2$ and 
${\rm Tr}\mathcal{M}_{P}^2$, which constrain the sum of the masses.

Although the singlet Higgs fields $h_{n}$ and $a_{n}$ do not couple to the gauge bosons, 
all the mass eigenstates of the neutral Higgs boson can interact with the 
$W$, $Z$ bosons and fermions, because the singlet fields are mixed with the doublet fields.
The couplings of the charged Higgs boson 
with the gauge bosons and quarks are identical to those in the MSSM.
In particular, the $VVh$-, $Zhh$- and $bbh$-vertices are important for the study of 
Higgs production and decay events in colliders\cite{pdg}.
At the LEP-type $e^+e^-$ collider, 
the dominant production processes of the neutral Higgs bosons are 
the Higgs strahlung process associating with $Z$ boson, the $W$-fusion production, and
the pair production processes if $m_{h_1}+m_{h_2}<\sqrt{s}$.
As for the decay of a light Higgs boson, $h\rightarrow b\bar{b}$ is the main mode, with subleading
$h\to\tau^{+}\tau^{-}$ mode. In both cases, the correction factors to the relevant Yukawa couplings
are the same with each other which characterize deviation from the MSM.
In addition to the processes, the gluon fusion and the Yukawa processes become 
important at high-energy hadron colliders\cite{hadron-collider}.

It is straightforward to read off those vertices from the kinetic terms of the 
Higgs bosons, and from the Yukawa coupling terms:
\begin{align}
\label{eq:coupling-hvv-mass-eigenstate}
 &\mathcal{L}_{VVh}=g_2m_Wg_{VVh_i}
\left(W_\mu^+ W^{-\mu}+\frac{1}{2\cos^2\th_W}Z_\mu Z^\mu\right)h_i,\\
\label{eq:coupling-hhz-mass-eigenstate}
 &\mathcal{L}_{Zhh}=
\frac{g_2}{2\cos\th_W}(h_d{\stackrel{\leftrightarrow}{\del}}_\mu a_d
-h_u{\stackrel{\leftrightarrow}{\del}}_\mu a_u)Z^\mu\\
\notag
 &\;\quad\quad=\frac{g_2}{2\cos\th_W}g_{Zh_ih_j}Z^\mu
(h_i{\stackrel{\leftrightarrow}{\del}}_\mu h_j),\\
\label{eq:lagrangian-hqq-dirac}
 &\mathcal{L}_{bbh}=-\frac{g_2m_b}{2m_W}\bar{b}
(g_{bbh_i}^S+i\g^5g_{bbh_i}^P)bh_i,
\end{align}
where the correction factors to the couplings are
\begin{align}
\label{eq:g-hvv}
 &g_{VVh_i}=O_{1i}\cos\b+O_{2i}\sin\b,\\
\label{eq:g-hhz}
 &g_{Zh_ih_j}=\half\left\{
(O_{4i}O_{2j}-O_{4j}O_{2i})\cos\b-(O_{4i}O_{1j}-O_{4j}O_{1i})\sin\b\right\},\\
\label{eq:g-hqq}
 &g_{bbh_i}^S=O_{1i}\frac{1}{\cos\b},\quad g_{bbh_i}^P=-O_{4i}\tan\b.
\end{align}
Then the mass eigenstates and the gauge eigenstates are related as  
$h_d=O_{1i}h_i$, $h_u=O_{2i}h_i$ and $a=O_{4i}h_i$, where $i$ is summed from $1$ to $5$.
The $Zh_ih_j$ coupling vanishes for $i=j$, because of the antisymmetric derivative nature of  its definition.
This coupling also vanishes when both the Higgs bosons are either scalars ($i,j=1,2,3$) or pseudoscalars ($i,j=4,5$).
However, all the $g_{Zh_ih_j}$ are expected to have nonzero values 
in the CP-violating case, because of the mixing of these CP eigenstates. 
The equations (\ref{eq:g-hvv})--(\ref{eq:g-hqq}) have the same structure as those in the MSSM\cite{squark-phase-mssm},
except for the mixing including the singlet.

\subsection{Constraints on the parameters}
\label{sec:constraints-parameter}
The NMSSM has more parameters than the MSSM.
Our main concern is to search for allowed parameters in the case of weak scale $v_n$, for which
we expect new features in the spectrum and coupling of the Higgs bosons, as well as the phase
transitions at finite temperature.
In order to select the allowed parameters, we impose the following two conditions on the model:
\begin{itemize}
\item[(1)] the vacuum condition, 
which requires that prescribed vacuum is the global minimum of the effective potential. 
This also requires that all the masses-squared of the scalars including the sfermions be positive.
\item[(2)] the spectrum condition, which requires that the mass of the Higgs boson with its couplings
to the vector boson $\left|g_{VVh}\right|$ larger than $0.1$ be heavier than the bound $114\mbox{GeV}$.
\end{itemize}
The constraint on the gauge coupling is the most stringent so that we examine the other couplings
for the allowed parameters later. 
Since the mass matrix of the Higgs bosons receives large radiative corrections, we need some numerical studies
to figure out the results of the spectrum condition, 
which will be presented in Section~\ref{sec:parameter-search}.
Here we attempt to find analytic form of constraints obtained from the vacuum condition at the tree level.

In the MSSM, the global minimum of the tree-level potential is always located at the vacuum, as long as
the $D$-flat direction is raised by the soft terms and the tadpole conditions are satisfied.
Although the Higgs potential in the NMSSM is bounded from below by the $F$-terms,
the prescribed vacuum is not always the global minimum of the potential, even when the tadpole conditions
are satisfied. This is because the trilinear terms in $\mathcal{L}_{\rm soft}$, which give negative contributions to
the potential, make some point different from the vacuum to be the global minimum.
We must exclude such a parameter set which yields an unwanted global minimum.
A necessary condition for the correct vacuum is that the mass-squared of all the scalars be positive.
In the CP-conserving case, this applied to the pseudoscalars implies that $\det{\mathcal{M}_P^{\prime}}^2>0$, hence,
\begin{equation}
\hat{m}^2(-\frac{3}{4}\mathcal{R}v^2\sin2\b+R_\k v_n)
> \frac{3}{4}\mathcal{R}^2v^2v_n^2.
\label{eq:charged-bound-lower}
\end{equation}
This requires that each factor in the left-hand side has the same sign, 
and gives the lower bound on the charged 
scalar mass, which is large enough for $\hat m^2>0$.

\begin{figure}[htbp]
\begin{center}
\includegraphics[height=6cm]{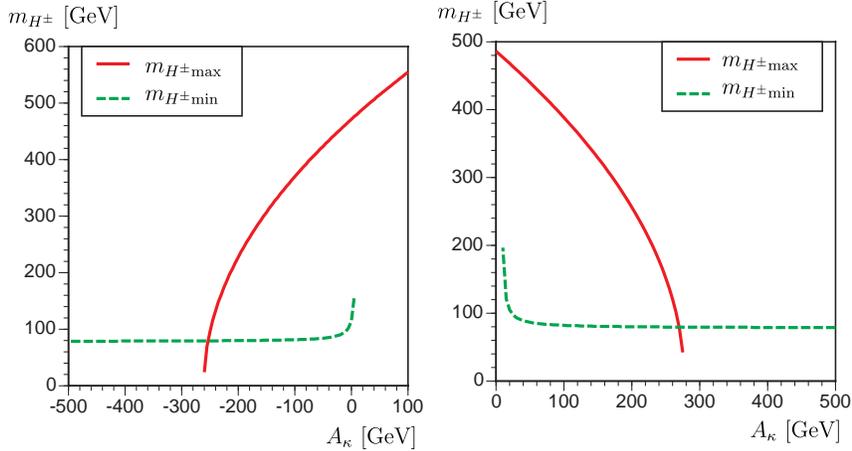}
\caption[charged Higgs bound] {
The mass bounds on the tree-level charged Higgs boson 
as a function of $A_\k$ for
$\tan\b=5$, $v_n=300{\rm GeV}$, $\lam=0.1$ and $\k=-0.3$ (left-hand plot)
and $\k=0.3$ (right-hand plot), respectively.
The solid line shows the upper mass bound of charged Higgs boson and
dashed one shows a lower bound.}
\label{fig:charged-higgs-bound}
\end{center}
\end{figure}

Another necessary condition is that the value of the scalar potential 
at prescribed vacuum be lower than that at the origin.
Now at the tree level, $V(0)$ is zero and that at the vacuum is
\begin{align}
\left.V\right|_{\rm vacuum}
 =&-\frac{1}{4}|\lam|^2v_n^2v^2-\frac{1}{4}|\k|^2v_n^4
-\frac{1}{8}m_Z^2v^2\cos^22\b-\frac{1}{8}m_W^2v^2\sin^22\b
\nonumber\\
 &+\frac{1}{8}m_{H^\pm}^2v^2\sin^22\b
  -\frac{1}{8}\mathcal{R}v_n^2v^2\sin2\b+\frac{1}{6}R_\k v_n^3,
\end{align}
where we use the tadpole conditions (\ref{eq:tad-hd})--(\ref{eq:tad-hn}) to eliminate the soft masses 
of the Higgs fields and use $m_{H^\pm}$ instead of $R_\lam$.
Then requiring that $\left.V\right|_{\rm vacuum}<0$ gives $m_{H^\pm}$ the upper bound
\begin{align}
 &m_{H^\pm}^2<
 2|\lam|^2v_n^2\frac{1}{\sin^22\b}+2|\k|^2\frac{v_n^4}{v^2}
 \frac{1}{\sin^22\b}+m_Z^2\cot^22\b+m_W^2
\nonumber\\ &\qquad\quad
 +\mathcal{R}v_n^2\frac{1}{\sin2\b}
 -\frac{4}{3}R_\k\frac{v_n^3}{v^2}\frac{1}{\sin^22\b}.
\label{eq:charged-bound-upper}
\end{align}
This bound is available regardless of whether there is CP violation or not.
The charged Higgs mass $m_{H^\pm}$ is not constrained in the MSSM-limit
where $\lam v_n$ and $\k v_n$ fixed for $v_{n} \rightarrow\infty$\cite{ellis89}, 
because of the infinitely large $v_n^4$ term. 
So this condition becomes important in the pure-NMSSM parameter set, 
i.e., when $v_n$ is not so large and $\lam$ and $\k$ are not so small.
Fig.~\ref{fig:charged-higgs-bound} show the tree-level charged 
Higgs mass bounds for an example of pure-NMSSM parameter set.
The solid line which shows the upper bound 
suggests that the charged Higgs boson must be lighter than $400\:{\rm GeV}$.
Hence the pure-NMSSM parameter set predicts the charged Higgs boson 
with the mass accessible to the LHC\cite{lhc-higgs}. 
The dashed line indicates the lower bound on the charged Higgs boson. 
It ends near $A_\k=0$ because the left hand side of (\ref{eq:charged-bound-lower})
becomes negative. 
The consistency of the model sets both the upper and lower bounds on $A_\k$,
which are of weak scale for a weak scale $v_n$.
In particular, the same sign of $\k$ as $A_\k$ is favored.

Once we fix $m_{H^\pm}$ and $A_\k$, the condition (\ref{eq:charged-bound-upper}) excludes
the elliptic region in the $(\lam , \k)$-plane, whose area vanishes in the MSSM-limit.
As we see below, this excludes a large portion of the parameter space for the case of
weak scale $v_n$.

\section{One-loop Effective Potential}
\label{sec:one-loop}
We now evaluate the one-loop contributions to the tadpole conditions and 
the spectrum of the Higgs bosons.
As we mentioned in the previous Section, the basic idea is almost the same 
as in the tree-level analysis, so we show the results briefly.
We analyze the effective potential of the Higgs fields by taking into account
the one-loop contributions from the gauge bosons 
and the third generation of the quarks and squarks. 
The corrections from the leptons, the other quarks and squarks can be ignored,
because of their small Yukawa couplings to the Higgs fields.
The corrections to the Higgs potential are given by
\begin{equation}
\label{eq:effective-potential}
\D V=\D_qV+\D_{\tilde{q}}V+\D_gV,
\end{equation}
where
\begin{align}
 \D_qV&=-\frac{N_C}{16\pi^2}\sum_{q=t,b}\left(\bar{m}_q^2\right)^2
 \left(\log\frac{\bar{m}_q^2}{M^2}-\frac{3}{2}\right),\\
 \D_{\tilde{q}}V&=\frac{N_C}{32\pi^2}\sum_{q=t,b}\sum_{j=1,2}
 \left(\bar{m}_{\tilde{q}_j}^2\right)^2
 \left(\log\frac{\bar{m}_{\tilde{q}_j}^2}{M^2}-\frac{3}{2}\right),\\
 \D_gV&=\frac{3}{64\pi^2}\left[\left(\bar{m}_Z^2\right)^2
 \left(\log\frac{\bar{m}_Z^2}{M^2}-\frac{3}{2}\right)
 +2\left(\bar{m}_W^2\right)^2
 \left(\log\frac{\bar{m}_W^2}{M^2}-\frac{3}{2}\right)\right].
\end{align}
The field-dependent masses $\bar{m}_X^2$ are listed 
in Appendix~\ref{sec:field-dependent-mass}.
The tadpole conditions and the mass matrix are obtained by calculating 
the derivatives of the effective potential at the vacuum with 
these corrections.
In particular, the imaginary parts of the tadpole conditions are
\begin{equation}
I_\lam+\D I_\lam=\half\mathcal{I}v_n
,\quad
I_\k=-\frac{3}{2}\mathcal{I}\frac{v_dv_u}{v_n},
\label{eq:tadpole-p-loop}
\end{equation}
where $I_\k$ does not receive the loop correction at this level.
The correction $\D I_\lam$ comes from the squark loops
\begin{equation}
\D I_\lam=\frac{N_C}{16\pi^2}\sum_{q=t,b}
|y_q|^2I_q\left[f(m_{\tilde{q}_1}^2,m_{\tilde{q}_2}^2)-1\right]v_n,
\end{equation}
where $I_q$ is defined by (\ref{eq:rq-iq}), and
\begin{equation}
 f(m_1^2,m_2^2)=\frac{1}{\D m^2}\left[
m_1^2\left(\log\frac{m_1^2}{M^2}-1\right)
-m_2^2\left(\log\frac{m_2^2}{M^2}-1\right)\right],
\end{equation}
with $\D m^2=m_1^2-m_2^2$. 

Here we show the corrections to the mass matrix in detail.
The couplings to the gauge bosons and the quarks 
receive the loop effects, through the orthogonal matrix $O$ which is 
determined from the corrections to the neutral mass matrix $\mathcal{M}$.
The scalar part $\mathcal{M}^2_S$ of the neutral mass matrix
receives the loop contributions,
\begin{equation}
 \D\mathcal{M}_S^2=\D_q\mathcal{M}_S^2+\D_g\mathcal{M}_S^2
 +\D_{\tilde{q}}\mathcal{M}_S^2,
\end{equation}
from the quarks, gauge bosons and squarks.
The quark loops and gauge loops contribute to the 
upper-left $2\times 2$ elements only,
\begin{equation}
\D_q\mathcal{M}_S^2
=-\frac{N_C}{4\pi^2}\left(\begin{array}{ccc}
|y_b|^2m_b^2\log\frac{m_b^2}{M^2}&0&0
\\
0&|y_t|^2m_t^2\log\frac{m_t^2}{M^2}&0
\\
0&0&0
\end{array}\right),
\end{equation}
\begin{equation}
 \D_g\mathcal{M}_S^2
=\frac{3}{32\pi^2}\left(m_Z^2\log\frac{m_Z^2}{M^2}
 +2m_W^2\log\frac{m_W^2}{M^2}\right)
\left(\begin{array}{ccc}
\cos^2\b&\cos\b\sin\b&0\\
\cos\b\sin\b&\sin^2\b&0\\
0&0&0\end{array}\right),
\end{equation}
because they do not couple to the singlet in the tree-level potential.
The squark-loop contributions are
\begin{align}
\label{eq:stop-correction-Ms}
 &\D_{\tilde{t}}\mathcal{M}_S^2=\frac{N_C}{16\pi^2}
\left[\mathcal{T}^Sf(m_{\tilde{t}_1}^2,m_{\tilde{t}_2}^2)
+\sum_{j=1,2}
 \left<\frac{\del\bar{m}_{\tilde{t}_j}^2}{\del \bm{h}}\right>
 \left<\frac{\del\bar{m}_{\tilde{t}_j}^2}{\del \bm{h}}\right>^T
 \log\frac{m_{\tilde{t}_j}^2}{M^2}
  \right],
\\
\label{eq:sbottom-correction-Ms}
 &\D_{\tilde{b}}\mathcal{M}_S^2=\frac{N_C}{16\pi^2}
\left[\mathcal{B}^Sf(m_{\tilde{b}_1}^2,m_{\tilde{b}_2}^2)
+\sum_{j=1,2}
 \left<\frac{\del\bar{m}_{\tilde{b}_j}^2}{\del \bm{h}}\right>
 \left<\frac{\del\bar{m}_{\tilde{b}_j}^2}{\del \bm{h}}\right>^T
 \log\frac{m_{\tilde{b}_j}^2}{M^2}
  \right],
\end{align}
where the matrices $\mathcal{T}$, $\mathcal{B}$ 
and the list of derivatives of the field dependent masses, 
are given in Appendix~\ref{sec:squark-derivatives}.

$\mathcal{M}_{P}^2$ and $\mathcal{M}_{SP}^2$ are not affected by
quark and gauge loops but receive the contributions from squark loops
\begin{equation}
 \D\mathcal{M}_P^2=\D_{\tilde{q}}\mathcal{M}_P^2
,\quad
 \D\mathcal{M}_{SP}^2=\D_{\tilde{q}}\mathcal{M}_{SP}^2,
\end{equation}
where squark-loop contributions are 
\begin{align}
\label{eq:stop-correction-Mp}
 &\D_{\tilde{t}}\mathcal{M}_P^2=\frac{N_C}{16\pi^2}
\left[\mathcal{T}^Pf(m_{\tilde{t}_1}^2,m_{\tilde{t}_2}^2)
+\sum_{j=1,2}
 \left<\frac{\del\bar{m}_{\tilde{t}_j}^2}{\del \bm{a}}\right>
 \left<\frac{\del\bar{m}_{\tilde{t}_j}^2}{\del \bm{a}}\right>^T
 \log\frac{m_{\tilde{t}_j}^2}{M^2}
  \right],
\\
\label{eq:sbottom-correction-Mp}
 &\D_{\tilde{b}}\mathcal{M}_P^2=\frac{N_C}{16\pi^2}
\left[\mathcal{B}^Pf(m_{\tilde{b}_1}^2,m_{\tilde{b}_2}^2)
+\sum_{j=1,2}
 \left<\frac{\del\bar{m}_{\tilde{b}_j}^2}{\del \bm{a}}\right>
 \left<\frac{\del\bar{m}_{\tilde{b}_j}^2}{\del \bm{a}}\right>^T
 \log\frac{m_{\tilde{b}_j}^2}{M^2}
  \right],
\end{align}
and
\begin{align}
\label{eq:stop-correction-Msp}
 &\D_{\tilde{t}}\mathcal{M}_{SP}^2=\frac{N_C}{16\pi^2}
\left[-\mathcal{T}^{SP}f(m_{\tilde{t}_1}^2,m_{\tilde{t}_2}^2)
+\sum_{j=1,2}
 \left<\frac{\del\bar{m}_{\tilde{t}_j}^2}{\del \bm{h}}\right>
 \left<\frac{\del\bar{m}_{\tilde{t}_j}^2}{\del \bm{a}}\right>^T
 \log\frac{m_{\tilde{t}_j}^2}{M^2}
  \right],
\\
\label{eq:sbottom-correction-Msp}
 &\D_{\tilde{b}}\mathcal{M}_{SP}^2=\frac{N_C}{16\pi^2}
\left[-\mathcal{B}^{SP}f(m_{\tilde{b}_1}^2,m_{\tilde{b}_2}^2)
+\sum_{j=1,2}
 \left<\frac{\del\bar{m}_{\tilde{b}_j}^2}{\del \bm{h}}\right>
 \left<\frac{\del\bar{m}_{\tilde{b}_j}^2}{\del \bm{a}}\right>^T
 \log\frac{m_{\tilde{b}_j}^2}{M^2}
  \right].
\end{align}
The NG mode can be extracted from $\mathcal{M}_{P}^2$ and 
$\mathcal{M}_{SP}^2$ 
by the same orthogonal transformation as at the tree level. 
The loop contributions to the off-diagonal matrix $\mathcal{M}_{SP}^2$ 
are proportional to the CP-violating parameters in the squark sector, 
$I_q$ which is defined in (\ref{eq:rq-iq}).

The charged Higgs mass is not affected by the singlet field 
but receives the contributions from the gauge, quark and squark loops,
\begin{equation}
\D m_{H^\pm}^2=\D_gm_{H^\pm}^2
+\D_qm_{H^\pm}^2+\D_{\tilde{q}}m_{H^\pm}^2,
\label{eq:charged-mass-loop}
\end{equation}
where the detailed form of the each terms are 
given in Appendix~C of our previous paper\cite{funakubo03}, 
except that $\mu$ in the MSSM
is replaced by $\lambda v_n e^{i\varphi}/\sqrt2$.
$R_{\lam}$, which is determined from equation (\ref{eq:R_lam}),
receives loop corrections through the corrections to the charged Higgs mass. 

\section{Parameter search}
\label{sec:parameter-search}
In this Section, we search for the allowed parameter region numerically with the one-loop corrections
in the CP-conserving case.
The allowed parameter sets are selected by requiring the two conditions discussed in
Section~\ref{sec:constraints-parameter}.
For this purpose, we scanned the parameters in the Higgs sector within the following region:
$\tan\b=3 - 20$, $v_n=100 - 1000\mbox{GeV}$, 
$100 \le m_{H^\pm}\le 5000$ GeV and $-1000\le A_\k\le 0$ GeV.
Since we can always make $\lam$ positive without loss of generality, $(\lam,\k)$-plane is
scanned for $0\le\lam \le 1$ and $-1\le \k\le 1$.
As for the squark sector, we adopt a small value of the $A$-term, $A_t=A_b=20\mbox{GeV}$,
for the squark fields not to acquire nonzero VEV, and three patterns of the soft masses:
the heavy-squark scenario with $(m_{\tilde{q}},\;m_{\tilde{t}_R})=(1000\mbox{GeV},\;800\mbox{GeV})$, 
the light-squark-1 with $(m_{\tilde{q}},\;m_{\tilde{t}_R})=(1000\mbox{GeV},\;10\mbox{GeV})$
and the light-squark-2 with $(m_{\tilde{q}},\;m_{\tilde{t}_R})=(500\mbox{GeV},\;10\mbox{GeV})$,
where $m_{\tilde{q}}$ ($m_{\tilde{t}_R}$) denotes the doublet (singlet) soft mass.
For simplicity, we put $m_{\tilde{b}_R}= m_{\tilde{t}_R}$.
First of all, we pick up a set of parameters except for $\lam$ and $\k$ and exclude the regions
in $(\lam,\k)$-plane where the effective potential at the origin is smaller than that at the vacuum
and the spectrum condition is not satisfied.
Within the remaining region, we perform the numerical search for the global minimum of the 
effective potential and exclude the points for which the minimum is different from the vacuum.

\begin{figure}
\centering
\includegraphics[height=6cm]{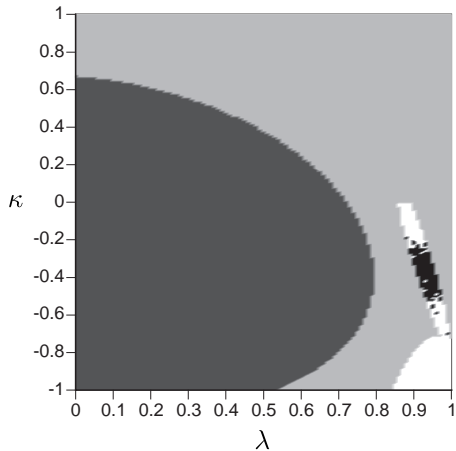}
\caption[A typical example]{
Allowed parameter region as a function of $\lam$ and $\k$ for 
$\tan\b=3$, $v_n=200$ GeV,
$m_{H^\pm}=400$ GeV and $A_\k=-200$ GeV for heavy-squark scenario.
Allowed parameter region are shown in white.}
\label{fig:lambda-kappa-allowed}
\end{figure}

In Fig.~\ref{fig:lambda-kappa-allowed}, 
we present a typical example of the allowed parameter region for $\tan\b=3$, $v_n=200$ GeV,
$m_{H^\pm}=400$ GeV and $A_\k=-200$ GeV for heavy-squark scenario.
The white region indicates the allowed parameter region in ($\lam$, $\k$)-plane.
Although negative $\k$ is favored for negative $A_\k$ 
as explained in Section~\ref{sec:constraints-parameter}, 
we study the range of $\k$ as $-1$ to $1$ since it is not excluded completely.
Within the dark gray elliptic region, the effective potential at the origin is smaller than that at the prescribed vacuum. 
This is predicted by (\ref{eq:charged-bound-upper}) which was derived from the tree-level potential.
The broad light gray region is excluded by the spectrum condition.
Within the upper allowed region near $\k=0$, the light Higgs scenario is realized, 
while the lower allowed region yields the lightest Higgs boson heavier than $114$ GeV.
The narrow black region right to the elliptic one is excluded, 
since the global minimum is located at a point different from the prescribed vacuum.
For these excluded parameters, the global minimum is located at
$v=0$ and a large $v_n$, which depends on $m_N^2$, $R_\k$ and $\left|\k\right|^2$.

\begin{figure}
\centering
\includegraphics[height=12.5cm]{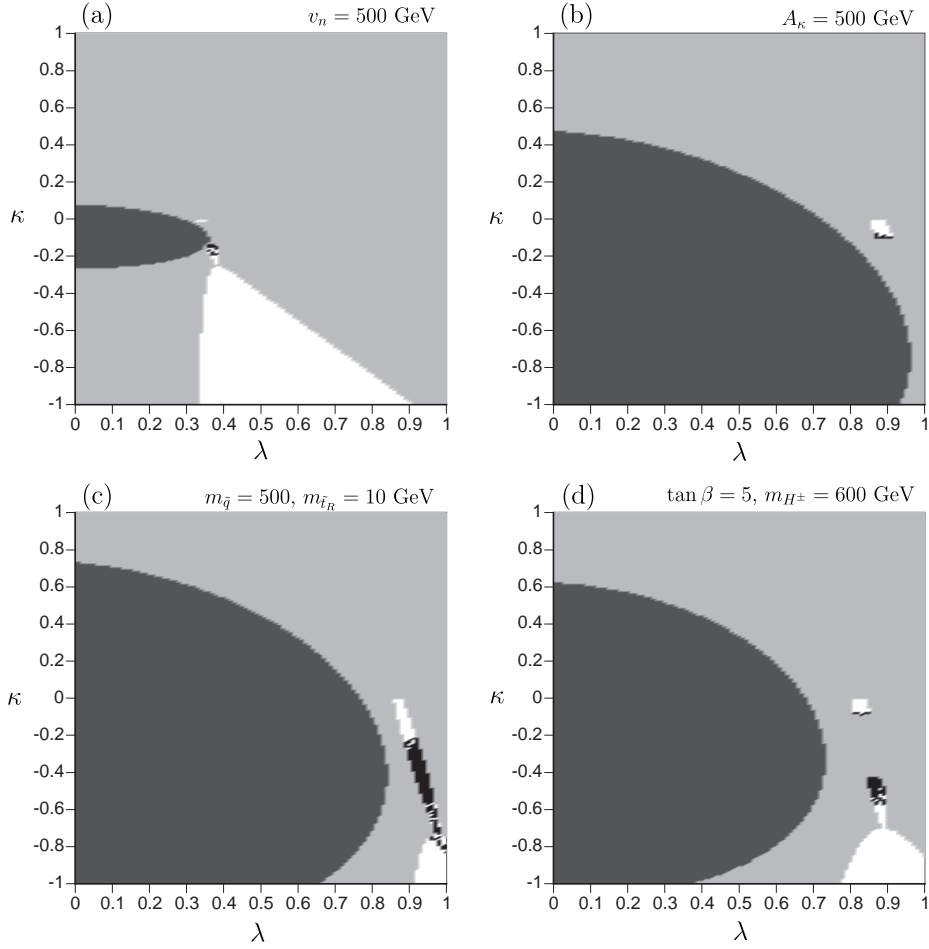}
\caption[parameter search]{
The same as Fig.~\ref{fig:lambda-kappa-allowed}, but with 
(a) $v_n=500$ GeV, 
(b) $A_\k$=500 GeV, 
(c) light-squark-2 scenario 
(i.e. $m_{\tilde{q}}=500$ GeV, $m_{\tilde{t}_R}=10$ and $A_t=20$ GeV) and 
(d) $\tan\b=5$ and $m_{H^\pm}=600$ GeV.}
\label{fig:parameter-search}
\end{figure}

We explain the dependence of the allowed regions on the parameters. 
In Fig.~\ref{fig:parameter-search}, 
we show the same as Fig.~\ref{fig:lambda-kappa-allowed}, but with 
(a) $v_n=500$ GeV, 
(b) $A_\k=500$ GeV, 
(c) light-squark-2 scenario and 
(d) $\tan\b=5$ and $m_{H^\pm}=600$ GeV.
As shown in Fig.~\ref{fig:parameter-search} (a), for a larger $v_n$,
the allowed region of the light Higgs shrinks and 
the allowed region of the heavy Higgs spreads out to small $\lam$ value.
The allowed parameter set with small $\lambda$ corresponds to the
MSSM limit. 
Although not shown in the graph, the similar behavior is observed, 
when the charged Higgs boson becomes heavier. 
For larger $A_\k$, the allowed region becomes smaller.
It is shown in Fig.~\ref{fig:parameter-search} (b) and was also expected from 
Fig.~\ref{fig:charged-higgs-bound}. 
If we choose small $m_{\tilde{q}}$ and $m_{\tilde{q}_R}$, 
Fig.~\ref{fig:parameter-search} (c) and the same plot for the 
light-squark-2 scenario show weak dependence on the squark soft masses.
Fig.~\ref{fig:parameter-search} (d) indicates that for larger $\tan\b$, 
the allowed region with light Higgs gets narrower.
At $\tan\b=20$, the allowed region becomes point like at $\k=0$, 
where one of the pseudoscalar is always  massless. 

\begin{figure}
\centering
\includegraphics[height=64mm]{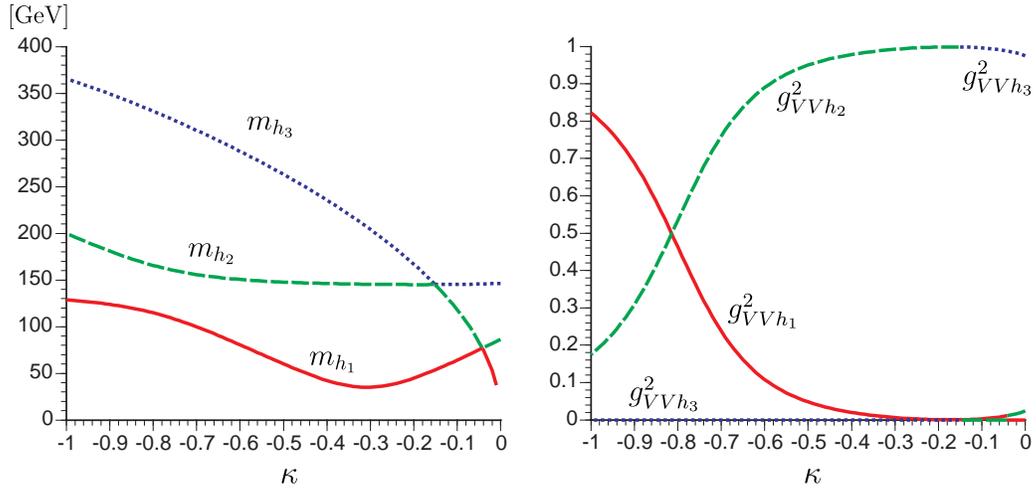}
\caption[mass and couplings] {
The Higgs masses (left) and the couplings (right) to the $V$-bosons
as a function of $\k$ for the same parameters as 
Fig.~\ref{fig:lambda-kappa-allowed} with $\lam=0.9$. 
In the left plot, solid line, broken line and dotted line 
represent $m_{h_1}$, $m_{h_2}$ and $m_{h_3}$, respectively. 
In the right plot, solid line, dashed line and dotted line represent 
$g_{VVh_1}^2$, $g_{VVh_2}^2$ and $g_{VVh_2}^2$, respectively. }
\label{fig:mass-coupling}
\end{figure}

\begin{figure}
\centering
\includegraphics[height=6cm]{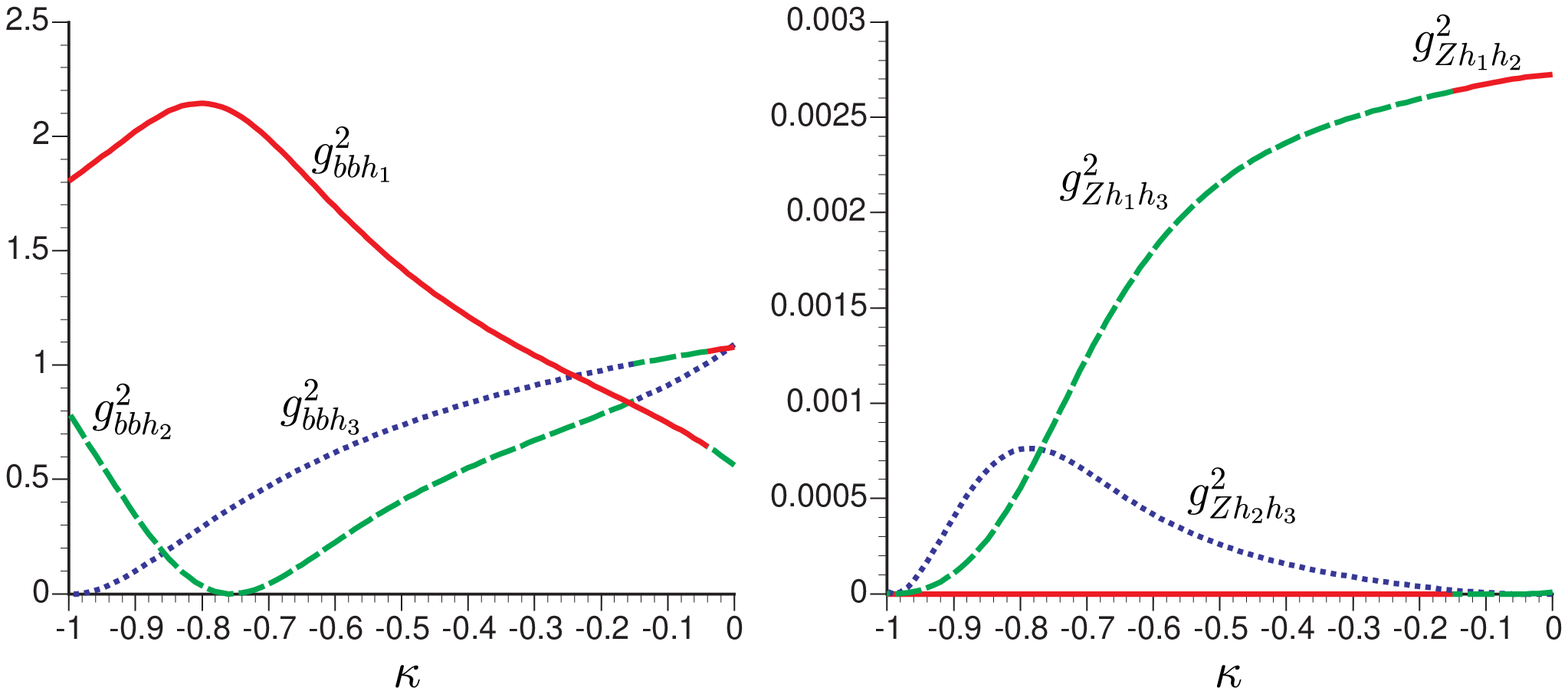}
\caption[couplings] {
The couplings of the three lightest Higgs bosons to the bottom quarks and 
the $Z$ boson as functions of $\kappa$ for the same parameters as 
Fig.~\ref{fig:mass-coupling}.
In the left plot, 
solid line, broken line and dotted line represent 
$g_{bbh_1}^2=(g_{bbh_1}^S)^2+(g_{bbh_1}^P)^2$, $g_{bbh_2}^2$, $g_{bbh_3}^2$, 
respectively.
In the right plot, solid line, broken line and dotted line represent 
$g_{Zh_1h_2}^2$, $g_{Zh_1h_3}^2$ and $g_{Zh_2h_3}^2$, respectively. }
\label{fig:couplings}
\end{figure}

Now, we consider the detail of the Higgs mass spectrum and couplings.
In Fig.~\ref{fig:mass-coupling} and \ref{fig:couplings}, we show the behaviors of the 
Higgs masses (Fig.~\ref{fig:mass-coupling} left) 
and the couplings of the three lightest Higgs bosons  
to the massive gauge bosons (Fig.~\ref{fig:mass-coupling} right), 
to the bottom quarks (Fig.~\ref{fig:couplings} left), 
and the $Zhh$-couplings (Fig.~\ref{fig:couplings} right), 
as a function of $\k$ for the same parameters as Fig.~\ref{fig:lambda-kappa-allowed} with $\lam=0.9$.
These correction factors to the coupling constants are defined in  (\ref{eq:g-hvv})--(\ref{eq:g-hqq}).
As seen from Fig.~\ref{fig:lambda-kappa-allowed}, 
the allowed region along $\lam=0.9$ is divided into two parts, 
one of which corresponds to the light Higgs scenario with small $|\k|$. 
The range of $\k$ satisfying the spectrum condition can be read from 
Fig.~\ref{fig:mass-coupling} to be $-0.33<\k<-0.05$, 
some part of which is excluded by the vacuum condition.
Within the other allowed region for $-1.0<\k<-0.8$, 
all the Higgs bosons are heavier than $114\:{\rm GeV}$. 
We refer to this parameter set as the heavy Higgs scenario.
In the scenario, the relatively heavy Higgs boson $h_{3}$ is almost decouple from the theory 
because its couplings $g_{VVh_{3}}$ and $g_{bbh_{3}}$ are small. 
The lighter two Higgs bosons are both CP-even scalars, 
hence in the scenario, 
the NMSSM behaves like the MSSM.
Particularly the lightest Higgs boson is heavier than $120\:{\rm GeV}$ and 
its coupling to the bottom quark is not so large, $g_{bbh_{1}}^{2}<2.2$, 
which escapes the experimental bounds as the MSSM does.
It is difficult to distinguish the NMSSM from the MSSM in the heavy Higgs scenario.

In the light Higgs scenario, the light Higgs bosons cannot 
be observed in the collider experiments.
The main processes for producing the Higgs bosons at the LEP~2 are 
the Higgs strahlung process and the $W$-fusion process, 
but both processes include small $VVh$-couplings 
(shown in the right plot of Fig.~\ref{fig:mass-coupling}) 
and hardly produce the light Higgs bosons.
Though the pair production becomes important when 
the total mass of the lightest scalar and the lightest pseudoscalar bosons 
is under the LEP~2 threshold, 
they cannot be created in pair\cite{2hdm-lep} because of so small coupling
$g_{Zh_1h_3}^2$ ($g_{Zh_1h_2}^2$) shown in the right plot of Fig.~\ref{fig:couplings}.
The Yukawa processes can be dominant where these processes are suppressed 
but the Yukawa processes cannot be observed unless the $bbh$-couplings  are 
fairly large\cite{yukawa-lep}.
In the left plot of Fig.~\ref{fig:couplings}, 
the Yukawa couplings $g_{bbh}$ are almost unity in the light Higgs region. 
With these couplings the light Higgs bosons $h_{1}$ and $h_{2}$ produced in the 
Yukawa processes are hardly observed. 
These arguments are essentially the same even if we consider the hadron colliders.
The SM-Higgs search experiments at the Tevatron 
mainly consider the strahlung process with $W$($Z$) 
in the low mass range $110<m_{h} <140\:{\rm GeV}$, 
and the results show that the signal efficiency is too small to find out the 
Higgs boson\cite{tevatron-result}.
At the LHC, the gluon fusion and the $W$-fusion processes become important.
Though the $h_{1}$ and $h_{2}$ production in $W$-fusion process is suppressed by their small 
gauge couplings, their production in gluon fusion process is not suppressed. 
However it is in general difficult to select the gluon fusion events from the backgrounds,
because the process produces the Higgs boson only.
Therefore $h_{3}$ is expected to be first detected in the collider experiment.
Then the model is very similar to the SM with the Higgs mass 
which can be as heavy as $150\mbox{GeV}$.

\section{Effects of the CP violation}
\label{sec:cp-violation}
Now we turn on CP violation and study its effects 
on the spectrum and gauge couplings of the Higgs bosons. 
Such an analysis in the MSSM by considering the CP violation 
in the squark sector was done in \cite{squark-phase-mssm}. 
Here we focus on the tree-level CP violation which does not exist in the MSSM
and is characterized by the parameter $\mathcal{I}$ 
defined in (\ref{eq:rr}).
In addition to $\mathcal{I}$, the complex parameters in the Higgs sector entering 
the mass-squared matrix are summarized in 
$\mathcal{R}$, $R_\lambda$ and $R_\kappa$.
When all the parameters are real as in the previous Section, 
one can freely assign their values.
However, if some of them are complex, 
we must arrange the parameters and the phases $\th$ and $\varphi$ 
to satisfy the tadpole conditions 
(\ref{eq:tadpole-p-loop}). 
Before presenting the numerical results, 
we explain how we parameterize the CP violation.

First of all, $R_\lam$ is fixed by the charged Higgs mass $m_{H^\pm}$ 
from the equations (\ref{eq:R_lam}) and (\ref{eq:charged-mass-loop}). 
The remainders, $\mathcal{R}$, $R_\k$ and $\mathcal{I}$, 
are determined by $\lam$, $\k$, $A_\k$, $\th$ and $\varphi$.
Here $\lam$, $\k$ and $A_\k$ are complex numbers 
and some of the phases are redundant.
We denote the phase of $\lam$, $\k$ and $A_\k$ as 
$\d_\lam$, $\d_\k$ and $\d_{A_\k}$, respectively.
Among those, independent phases are corrected into 
$\d_\k^\prime\equiv\d_\k+3\varphi$, $\d_{A_\k}$ and 
$\d_{\rm EDM}\equiv\d_\lam+\th+\varphi$, which is effective to nEDM if the gaugino 
masses and $A_q$ are real.
The counterpart of the $\d_{\rm EDM}$ in the MSSM 
is the phase of the $\mu$-term plus $\th$. 
Suppose that we first give $|\lam|$, $|\k|$, $\d_{\rm EDM}$ and 
$\d_\k^\prime$, from which $\mathcal{R}$ and $\mathcal{I}$ are determined.
Since $I_{\k}$ is fixed by (\ref{eq:tadpole-p-loop}), 
one can take any absolute value of $A_{\k}$ but not its phase.
In particular, $R_{\k}$ is given, without specifying $\d_{A_{\k}}$, by the relation 
\begin{equation}
 R_\k^2={1\over2}|\k A_\k|^2-I_\k^2
 = {1\over2}|\k A_\k|^2 
 -\left({{3\mathcal{I}v_dv_u}\over{2v_n}}\right)^2,
\label{eq:Rk-fix-Ak}
\end{equation}
where we use the second equation of (\ref{eq:tadpole-p-loop}).
Not that, the $|\k A_\k|$ must be larger than 
$3\mathcal{I}v_dv_u/\sqrt{2}v_n$ in order for $R_\k$ to have a value.
In the following, we adopt the phase convention 
so that the relevant phase to the nEDM vanishes. 
Therefore, $\d'_{\k}$ is the only phase that we can freely assign its value.

\begin{figure}[ht]
\begin{center}
\includegraphics[height=64mm]{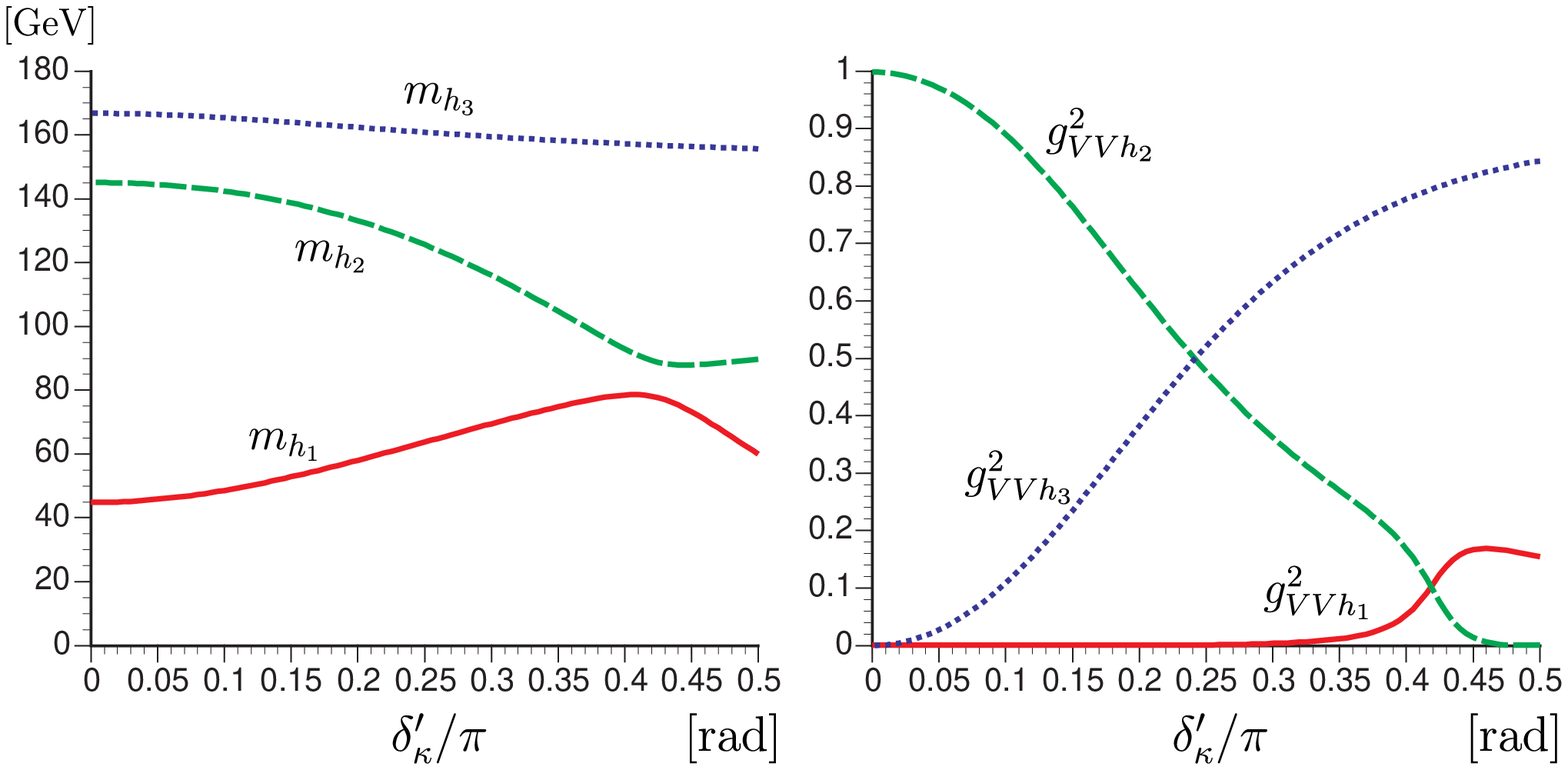}
\caption[The mass and coupling as a function of CP phase at the light Higgs]{
The same as Fig.~\ref{fig:mass-coupling}
but with $\k=-0.2$ (light Higgs scenario) and $\d_{\rm EDM}=0$ as a function of $\d_\k^\prime$.}
\label{fig:mass-coupling-cpv-LH}
\end{center}
\end{figure}

\begin{figure}[ht]
\begin{center}
\includegraphics[height=64mm]{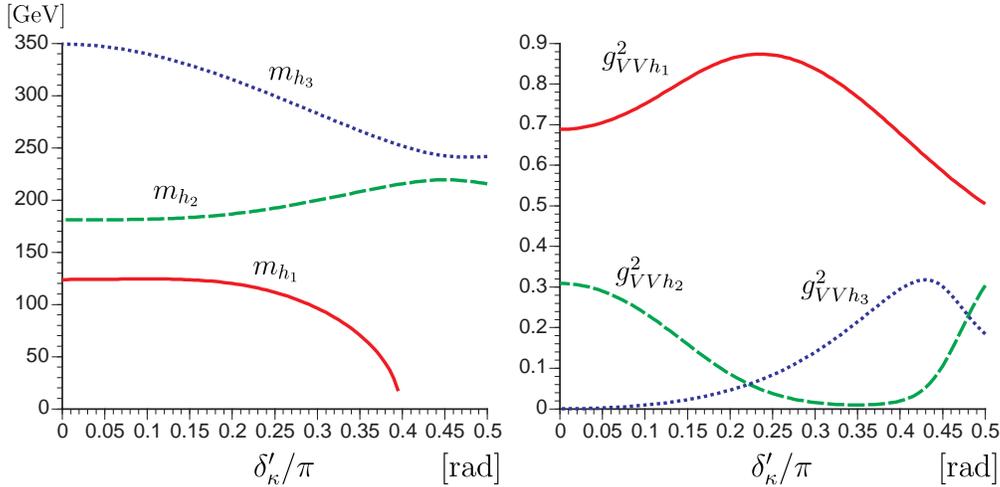}
\caption[The mass and coupling as a function of CP phase at the heavy Higgs]{
The same as Fig.~\ref{fig:mass-coupling-cpv-LH}
but with $\k=-0.9$ (heavy Higgs scenario).}
\label{fig:mass-coupling-cpv-HH}
\end{center}
\end{figure}

Finally we show the effects of $\delta'_\kappa$ on the masses and couplings of the
Higgs mass eigenstates with $\d_{\rm EDM}=0$.
For illustration, we take two parameter sets 
with $\k=-0.2$ and $\k=-0.9$, 
while the other parameters are the same as in Fig.~\ref{fig:mass-coupling}, 
and plot $\delta'_\k$-dependences of the masses and couplings 
in Fig.~\ref{fig:mass-coupling-cpv-LH} and Fig.~\ref{fig:mass-coupling-cpv-HH}, respectively.
In order to compare with the former result, 
positive $R_\k$ in the equation (\ref{eq:Rk-fix-Ak}) was chosen.
As shown in Fig.~\ref{fig:mass-coupling-cpv-LH} and Fig.~\ref{fig:mass-coupling-cpv-HH}, 
the phase dependences in the light Higgs scenario ($\k=-0.2$) are smaller  than 
in the heavy Higgs scenario ($\k=-0.9$), 
since $\left|\kappa\right|$ is smaller for the light Higgs scenario.
In Fig.~\ref{fig:mass-coupling-cpv-LH}, the next-to-lightest Higgs boson $h_{2}$ becomes lighter than 
$114\:{\rm GeV}$ for $\delta'_\kappa/\pi < 0.31$, hence the large $\d'_{k}$ is not allowed 
by the spectrum condition.
In Fig.~\ref{fig:mass-coupling-cpv-HH}, 
the model is excluded for $\d_{\k}^{\prime}/\pi>0.24$, because of 
the small Higgs mass with moderate gauge coupling 
and also excluded for $\d_{\k}^{\prime}/\pi>0.39$ where the prescribed vacuum becomes unstable.
We could not find an example in which an excluded parameter set is converted to be 
allowed as a light Higgs by introducing the CP violation in the tree-level Higgs sector. 
However the light Higgs boson is realized even for the heavy Higgs parameter sets 
when we introduce CP violation in the squark sector, 
in the same way as in the MSSM.

\section{Summary}
\label{sec:summary}
We investigated the spectrum and coupling constants of the Higgs bosons in the 
NMSSM, when the vacuum expectation value of 
the singlet is of the weak scale for a wide range of all the parameters in the model.
We formulated the mass matrices of the neutral Higgs bosons and the charged Higgs mass
in a manner independent of the phase convention in CP-violating case.
By use of the effective potential including one-loop corrections of the third generation
of quarks and squarks, we obtained constraints on the parameters of the model.
Then we required that the neutral Higgs boson whose coupling to the $Z$ boson is
not so small must be heavier than $114\mbox{GeV}$, and that
the prescribed vacuum must be the absolute minimum of the effective potential.
The latter is nontrivial in the NMSSM, in contrast to the MSSM, in which the electroweak vacuum is
the global minimum of the potential as long as the symmetry breaking conditions are satisfied.
We found that the vacuum condition leads to an upper bound on the charged Higgs boson, 
which is irrelevant in the MSSM limit but effective in our case of $\langle N\rangle=O(100)\mbox{GeV}$.
The allowed parameter regions are classified into two distinct sets: one admits a light scalar
with very small gauge coupling, while the other contains the scalars heavier than the bound.
The former is realized only in the case of weak scale $\langle N\rangle$ for $\kappa\simeq 0$.
In this light Higgs scenario, 
the light Higgs bosons are not excluded by currently published experimental data: 
the correction factors to the Yukawa coupling of the light bosons are not so large 
and those to the $Zhh$-coupling are so small that 
the light bosons have not been observed yet by experiments. 
Therefore the lightest scalar among those with large gauge coupling, 
which is the candidate produced in future collider experiments, 
behaves just like the Higgs boson in the standard model.
The SM-like Higgs boson can be as heavy as $150\mbox{GeV}$, 
so that the model will take place of the MSSM
if the first observed Higgs boson is heavier than $135\mbox{GeV}$.

Another feature of the Higgs sector in the NMSSM is a possible CP violation at the tree level.
We studied an explicit CP violation in the Higgs sector, which does not affect the neutron
EDM. Such a CP violation is expected to play an important role in the scenario of
electroweak baryogenesis. 
For several sets of allowed parameters in the CP-conserving case, we gradually introduced
a CP phase and studied its effects on the masses and couplings of the Higgs bosons.
As expected, the effect is larger for the parameter sets in the heavy Higgs region,
which has larger $\left|\kappa\right|$ than the light Higgs scenario.
We, however, could not find a case that a large CP violation makes the lightest Higgs boson
lighter than the present bound on the SM Higgs, while making its gauge coupling tiny not to
be produced in lepton colliders.
Such a situation has been observed in the MSSM, in which the CP violation in the squark sector
induces a large CP violation in the Higgs sector.
If such a large CP violation is common to all the generation of squarks, it should be
constrained by the neutron EDM experiments.
We have found that such a CP violation weakens 
the first-order EWPT in the MSSM with a light stop\cite{funakubo03}. 
Thus the available parameter region for electroweak baryogenesis is very limited in the MSSM.
As naively expected, if the EWPT in the NMSSM is strongly first order because of the trilinear terms 
in the Higgs potential, it will open new possibility of the baryogenesis\cite{nmssm-ewpt}. 
Since the strong phase transition will not be caused by a light stop, it will persist in the presence 
of the CP violation in the Higgs sector.
Further, the CP violation is not so strongly constrained by the EDM experiments.
A study of the EWPT in the NMSSM is now in progress.

\section*{Acknowledgements}
\label{sec:acknowledgements}
The authors gratefully thank F.~Toyoda, A.~Kakuto and S.~Otsuki 
for valuable discussions.
This work was supported by the kakenhi of the MEXT, Japan, No.~13135222.

\appendix
\section{The field dependent masses}
\label{sec:field-dependent-mass}
In this appendix, we summarize the field-dependent masses of the quarks
and squarks of the third generation and those of the gauge bosons.
We  retain only the neutral components of the Higgs fields, which  appear in
the definition of the effective potential (\ref{eq:effective-potential}) and
are necessary for the neutral Higgs mass-squared matrix.
For the charged Higgs mass, we need the expressions including the charged
Higgs fields $\phi_d^-$ and $\phi_u^+$, but they are almost the same as those
in the MSSM\cite{funakubo03} except for replacement of $\mu$ with
$\lambda v_n e^{i\varphi}/\sqrt2$.

The quark masses are expressed in terms of the 
Higgs fields and vacuum expectation values, 
\begin{align}
 &\bar{m}_b^2=|y_b|^2|\phi_d^0|^2
 =\half|y_b|^2(v_d^2+2v_dh_d+h_d^2+a_d^2)
,\\
 &\bar{m}_t^2=|y_t|^2|\phi_u^0|^2
 =\half|y_t|^2(v_u^2+2v_uh_u+h_u^2+a_u^2),
\end{align}
where $\phi_d^0=(v_d+h_d+ia_d)/\sqrt2$ and 
$\phi_u^0=e^{i\theta}(v_u+h_u+ia_u)/\sqrt2$.
If we take the Higgs fields as $h_d=h_u=h_n=0$, 
the masses at the vacuum are obtained, 
\begin{equation}
\left<\bar{m}^2_b\right>=m_b^2=\half|y_b|^2v_d^2,\quad
\left<\bar{m}^2_t\right>=m_t^2=\half|y_t|^2v_u^2,
\end{equation}
where the angle brackets denote vacuum expectation value.
The field dependent masses of the gauge bosons are
\begin{equation}
\bar{m}_Z^2=\frac{g_2^2+g_1^2}{2}(|\phi_d^0|^2+|\phi_u^0|^2)
,\quad
\bar{m}_W^2=\frac{g_2^2}{2}(|\phi_d^0|^2+|\phi_u^0|^2).
\end{equation}
Then the gauge bosons masses are
\begin{equation}
\left<\bar{m}_Z^2\right>=m_Z^2=\frac{g_2^2+g_1^2}{4}(v_d^2+v_u^2)
,\quad
\left<\bar{m}_W^2\right>=m_W^2=\frac{g_2^2}{4}(v_d^2+v_u^2).
\end{equation}
Similarly, field dependent top- and bottom-squark masses are
\begin{align}
 &\bar{m}_{\tilde{t}_{1,2}}^2
 =\half\left[m_{\tilde{q}}^2+m_{\tilde{t}_R}^2
 +\frac{g_2^2+g_1^2}{4}(|\phi_d^0|^2-|\phi_u^0|^2)
 +2|y_t|^2|\phi_u^0|^2\right.
\nonumber\\ &\qquad\qquad\quad
\left.\pm\sqrt{(m_{\tilde{q}}^2-m_{\tilde{t}_R}^2
 +x_t(|\phi_d^0|^2-|\phi_u^0|^2))^2
 +4|y_t|^2|\lam n\phi_d^0-A_t^*\phi_u^{0*}|^2}\:\right],
\\ &\bar{m}_{\tilde{b}_{1,2}}^2
 =\half\left[m_{\tilde{q}}^2+m_{\tilde{b}_R}^2
 -\frac{g_2^2+g_1^2}{4}(|\phi_d^0|^2-|\phi_u^0|^2)
 +2|y_b|^2|\phi_d^0|^2\right.
\nonumber\\ &\qquad\qquad\quad
\left.\pm\sqrt{(m_{\tilde{q}}^2-m_{\tilde{b}_R}^2
 +x_b(|\phi_d^0|^2-|\phi_u^0|^2))^2
 +4|y_b|^2|\lam n\phi_u^0-A_b^*\phi_d^{0*}|^2}\:\right],
\end{align}
where
\begin{equation*}
x_t\equiv\frac{1}{4}\left(g_2^2-\frac{5}{3}g_1^2\right),\quad
x_b\equiv-\frac{1}{4}\left(g_2^2-\frac{1}{3}g_2^2\right).
\end{equation*}
The masses of the squarks at the vacuum are
\begin{align}
 &\left<\bar{m}_{\tilde{t}_{1,2}}^2\right>=m_{\tilde{t}_{1,2}}^2=
  \half\left[m_{\tilde{q}}^2+m_{\tilde{t}_R}^2
  +\frac{g_2^2+g_1^2}{8}(v_d^2-v_u^2)+|y_t|^2v_u^2\right.
\nonumber\\ &\qquad\qquad\qquad\qquad\qquad\qquad\qquad\qquad\quad
  \left.\pm\sqrt{M_t^2+2|y_t|^2(P_tv_d^2+Q_tv_u^2)}\right],
\\
 &\left<\bar{m}_{\tilde{b}_{1,2}}^2\right>=m_{\tilde{b}_{1,2}}^2=
  \half\left[m_{\tilde{q}}^2+m_{\tilde{b}_R}^2
  -\frac{g_2^2+g_1^2}{8}(v_d^2-v_u^2)+|y_b|^2v_d^2\right.
\nonumber \\ &\qquad\qquad\qquad\qquad\qquad\qquad\qquad\qquad\quad
  \left.\pm\sqrt{M_b^2+2|y_b|^2(P_bv_u^2+Q_bv_d^2)}\right],
\end{align}
where we define the following combinations of the parameters:
\begin{align}
 \label{eq:rq-iq}
 &R_q=\frac{1}{\sqrt{2}}{\rm Re}(\lam A_qe^{i(\th+\phi)}),&
 &I_q=\frac{1}{\sqrt{2}}{\rm Im}(\lam A_qe^{i(\th+\phi)}),\quad (q=t,b),
\notag \\
 &P_t=\half|\lam|^2v_n^2-R_tv_n\tan\b,&
 &Q_t=|A_t|^2-R_tv_n\cot\b,
\notag \\%
 &P_b=\half|\lam|^2v_n^2-R_bv_n\cot\b,&
 &Q_b=|A_b|^2-R_bv_n\tan\b,
\notag \\%
 &M_t^2=m_{\tilde{q}}^2-m_{\tilde{t}_R}^2+\frac{x_t}{2}(v_d^2-v_u^2),&
 &M_b^2=m_{\tilde{q}}^2-m_{\tilde{b}_R}^2+\frac{x_b}{2}(v_d^2-v_u^2).
\end{align}
Although $I_q$ doesn't appear above equations, 
we defined this for later convenience 
(We use $I_q$ in Appendix \ref{sec:squark-derivatives}).

\section{Derivatives of the squark masses}
\label{sec:squark-derivatives}
The corrections from the squark loops to the neutral-mass matrix 
contain first derivatives of the field-dependent-squark masses and 
the matrices $\mathcal{T}^S$, $\mathcal{T}^P$, $\mathcal{T}^{SP}$, 
$\mathcal{B}^S$, $\mathcal{B}^P$ and $\mathcal{B}^{SP}$ 
(\ref{eq:stop-correction-Ms})--(\ref{eq:sbottom-correction-Msp}).
The First derivatives of the squark masses are as follows, 
\begin{align}
 &\left<\frac{\del\bar{m}_{\tilde{t}_{1,2}}^2}{\del \bm{h}}\right>
 =\left(\begin{array}{c}
 \frac{g_2^2+g_1^2}{8}v_d\\\left(|y_t|^2-\frac{g_2^2+g_1^2}{8}\right)v_u\\0
 \end{array}\right)
 \pm\frac{\bm{t}}{2\D m_{\tilde{t}}^2}
,\quad
 \left<\frac{\del\bar{m}_{\tilde{t}_{1,2}}^2}{\del \bm{a}}\right>
 =\pm\frac{|y_t|^2}{\D m_{\tilde{t}}^2}I_tv_n\bm{p},
\\%
 &\left<\frac{\del\bar{m}_{\tilde{b}_{1,2}}^2}{\del \bm{h}}\right>
 =\left(\begin{array}{c}
 \left(|y_b|^2-\frac{g_2^2+g_1^2}{8}\right)v_d\\\frac{g_2^2+g_1^2}{8}v_u\\0
 \end{array}\right)
 \pm\frac{\bm{b}}{2\D m_{\tilde{b}}^2}
,\quad
 \left<\frac{\del\bar{m}_{\tilde{b}_{1,2}}^2}{\del \bm{a}}\right>
 =\pm\frac{|y_b|^2}{\D m_{\tilde{b}}^2}I_bv_n\bm{p},
\end{align}
where
\begin{align}
 &\bm{t}=\left(\begin{array}{c}
 (x_tM_t^2+2|y_t|^2P_t)v_d\\(-x_tM_t^2+2|y_t|^2Q_t)v_u\\
 2|y_t|^2P_t\frac{v_d^2}{v_n}
 \end{array}\right),&
 &\bm{b}=\left(\begin{array}{c}
 (x_bM_b^2+2|y_b|^2Q_b)v_d\\(-x_bM_b^2+2|y_b|^2P_b)v_u\\
 2|y_b|^2P_b\frac{v_u^2}{v_n}
 \end{array}\right),
\notag \\%
 &\bm{p}=\frac{1}{v_n}
  \left(\begin{array}{c}v_uv_n\\v_nv_d\\v_dv_u\end{array}\right),&
 &\D m_{\tilde{q}}^2=m_{\tilde{q}_1}^2-m_{\tilde{q}_2}^2,\quad (q=t,b).
\end{align}
The detailed form of the matrices $\mathcal{T}^S$ and $\mathcal{B}^S$ are,
\begin{equation}
\mathcal{T}^S=\left(\begin{array}{ccc}
\frac{x_t^2}{2}v_d^2&-\frac{x_t^2}{2}v_dv_u&|y_t\lam|^2v_nv_d\\
-\frac{x_t^2}{2}v_dv_u&\frac{x_t^2}{2}v_u^2&0\\
|y_t\lam|^2v_nv_d&0&0
\end{array}\right)
+|y_t|^2R_t\left(\begin{array}{ccc}
\frac{v_uv_n}{v_d}&-v_n&-v_u\\
-v_n&\frac{v_nv_d}{v_u}&-v_d\\
-v_u&-v_d&\frac{v_dv_u}{v_n}
\end{array}\right)
-\frac{\bm{t}\bm{t}^T}{2(\D m_{\tilde{t}}^2)^2},
\end{equation}
and
\begin{equation}
\mathcal{B}^S=\left(\begin{array}{ccc}
\frac{x_b^2}{2}v_d^2&-\frac{x_b^2}{2}v_dv_u&0\\
-\frac{x_b^2}{2}v_dv_u&\frac{x_b^2}{2}v_u^2&|y_b\lam|^2v_nv_u\\
0&|y_b\lam|^2v_nv_u&0
\end{array}\right)
+|y_b|^2R_b\left(\begin{array}{ccc}
\frac{v_uv_n}{v_d}&-v_n&-v_u\\
-v_n&\frac{v_nv_d}{v_u}&-v_d\\
-v_u&-v_d&\frac{v_dv_u}{v_n}
\end{array}\right)
-\frac{\bm{b}\bm{b}^T}{2(\D m_{\tilde{b}}^2)^2}.
\end{equation}
These are obtained from the second derivatives of the $\D_{\tilde{q}}V$  
with respect to $h_d$, $h_u$ and $h_n$, 
and applying the tadpole conditions. 
The corrections to the pseudoscalar 
and scalar-pseudoscalar-mixing 
component of the neutral-mass matrix 
include following matrices: 
\begin{equation}
\mathcal{T}^P
=\left[|y_t|^2R_t\frac{v_n}{v_dv_u}
-2\left(\frac{|y_t|^2I_tv_n}{\D m_{\tilde{t}}^2}\right)^2\right]
\bm{p}\bm{p}^T
,\quad 
\mathcal{B}^P
=\left[|y_b|^2R_b\frac{v_n}{v_dv_u}
-2\left(\frac{|y_b|^2I_bv_n}{\D m_{\tilde{b}}^2}\right)^2\right]
\bm{p}\bm{p}^T.
\end{equation}
and
\begin{equation}
\mathcal{T}^{SP}
=\frac{|y_t|^2I_tv_n}{(\D m_{\tilde{t}}^2)^2}\bm{t}\bm{p}^T,
\quad
\mathcal{B}^{SP}
=\frac{|y_b|^2I_bv_n}{(\D m_{\tilde{b}}^2)^2}\bm{b}\bm{p}^T.
\end{equation}


\end{document}